\documentclass[prc,showpacs,12pt]{revtex4}
\usepackage{graphicx}
\usepackage{dcolumn}
\usepackage{bm}
\usepackage{longtable}

\begin{document}

\title{Mass distributions for induced fission of different Hg isotopes}
\author{A.V.Andreev, G.G.Adamian, N.V.Antonenko}
\affiliation{
Joint Institute for Nuclear Research, 141980 Dubna, Russia
}

\begin{abstract}
With the improved scission-point model the mass distributions are calculated
for induced fission of different Hg isotopes with the masses 180--196.
The drastic change in the shape of the mass distribution
from asymmetric to symmetric is revealed with increasing mass number of the fissioning Hg isotope,
and the reactions are proposed to verify this prediction experimentally.
The asymmetric mass distribution of
fission fragments observed in the recent experiment on the fission of $^{180}$Hg
is explained. The calculated mass distribution
and mean total kinetic energy of fission fragments
are in a good agreement with the available experimental data.

\end{abstract}

\pacs{24.75.+i, 25.85.-w, 24.10.Pa \\ Key words:
Binary fission; Mass distribution; Scission point model; Dinuclear system model}

\maketitle


Since the discovery of nuclear fission this phenomenon
is intensively investigated.
The mass distributions in the low-energy fission
of the nuclei of actinide region were explored in details.
The asymmetric shape of the mass distribution is well known
in the spontaneous, neutron induced, and $\beta$-delayed
fission of most actinide isotopes.
Such asymmetric
shape was theoretically explained by taking into account
the shell structure of the fragments~\cite{goenn,hall,oganess}. At the last decades
new experimental techniques were developed to be able
to investigate the low-energy fission of lighter isotopes.
The studies of Coulomb-excited fission of radioactive beams~\cite{schmidt}
revealed the predominance of symmetric fission in the light
thorium to astatine region. In the fission of stable targets
with masses 185--210 induced by bombardment with protons and $^{3,4}$He~\cite{itkis}
the mass distribution was also found to be symmetric in most cases.
However, for several nuclei with the masses about 200 the mass
distribution looks symmetric but with a small dip at the top.
Based on the most of experimental data, 
one could conclude that the asymmetric shape of mass distribution
in low-energy fission changes to symmetric with decreasing
mass number of the fissioning nucleus. It was unexpected
that in the recent experiment~\cite{andre}
on $\beta$-delayed fission of $^{180}$Tl (the fissioning
nucleus is $^{180}$Hg) the shape of the mass distribution was
found to be clearly asymmetric. The explanation of this
outstanding result is a challenge for nuclear theory and
a good test for the existing models of nuclear fission.

As shown in Refs.~\cite{sp,rand}, the observable characteristics of fission process are formed near
the prescission configurations of fissioning nucleus. Indeed, with the modified
scission-point model~\cite{and} one can describe the experimental data on fission
of actinides:
mass, charge, and kinetic energy distributions, neutron
multiplicity distributions. A new explanation of a
bimodality effect in fission of heavy actinides and fine structure of
mass-energy distribution in fission of $^{236}$U have been proposed.
Our model have been also
extended to the description of ternary fission~\cite{and1}.
The advantage of our model is that it allows us
to describe various experimental data with the fixed set
of parameters and assumptions. The wide range
of described fission observables and effects demonstrates the
predictive power of the model. In the present work we
apply our model to the fission of lighter nuclei for describing
the new experimental data on asymmetric fission of $^{180}$Hg.

Here, we give a short description
of the model, the details can be found in Refs.~\cite{and,and1}. 
The fissioning nucleus at the scission point is modeled
by the two nearly touching coaxial spheroids ---
fragments of a dinuclear system with the masses $A_L$, $A_H$ and
charges $Z_L$, $Z_H$ of the light ($L$) and heavy ($H$) fragments, respectively.
$A=A_L+A_H$ and $Z=Z_L+Z_H$ are the mass and charge numbers of
a fissioning nucleus, respectively.
Taking into account the volume conservation, the
shape of the system is defined by the mass and charge numbers
of the fragments, deformation parameters of the fragments $\beta_i$, and 
interfragment distance $R$.
The deformation parameter of each fragment
is the ratio of the major and minor semi-axes of the spheroid
$\beta_i=c_i/a_i$. Here and further, $i=L,H$ denotes the light
and heavy fragments of the dinuclear system. The case $\beta=1$ corresponds
to the spherical shape of the fragment. For small values of $\beta$,
the following equality is valid: $\beta\approx\beta_2+1$, where
$\beta_2$ is the parameter of quadrupole deformation in the
multiple expansion of the fragment shape.

The potential energy of the system is the sum of the
liquid drop energies $U^{LD}_i$ of each fragment, the shell correction
terms $\delta U^{sh}_i$, the energy of interaction of
the fragments $V^C + V^N$, and the rotational energy $V^{rot}$.
The shell corrections are calculated with the
Strutinsky method and two-center shell model~\cite{tcsh}, the damping of the
shell corrections with excitation is introduced in our model.
The interaction energy consists of the Coulomb interaction
of two uniformly charged spheroids and nuclear interaction
in the form of a double folding of nuclear densities and
density-dependent Skyrme-type nucleon-nucleon forces~\cite{poten}.
\begin{eqnarray}
U(\{A_i,Z_i,\beta_i\},R)&=&U^{LD}_L(A_L,Z_L,\beta_L)+
U^{LD}_H(A_H,Z_H,\beta_H)\nonumber\\
&+&\delta U^{sh}_L(A_L,Z_L,\beta_L)+
\delta U^{sh}_H(A_H,Z_H,\beta_H)\nonumber\\
&+& V^C(\{A_i,Z_i,\beta_i\},R) + V^N(\{A_i,Z_i,\beta_i\},R)\nonumber\\
&+& V^{rot}(\{A_i,Z_i,\beta_i\},R,l).
\label{ener}
\end{eqnarray}
All these terms, including the shell correction terms, depend on
deformations of the fragments.
For given deformations of the fragments the nuclear interaction
potential has a potential minimum (pocket) as a function
of the interfragment distance $R$. For calculation of the potential
energy we take the value of interfragment distance corresponding
to this minimum. Depending on the masses of the fragments and
their deformations the calculated distance between the tips of
the spheroids is 0.5--1~fm. 

The thermodynamical equilibrium is postulated at the scission point.
The excitation energy of the nuclear system at scission is calculated
as a difference between the potential energy $U_{g.s.}$ of the
fissioning nucleus in the ground state and the
potential energy $U$ of dinuclear system at the scission point plus the
initial excitation energy of the fissioning nucleus:
$E^*=U_{g.s.}-U+E^*_{CN}$.
The temperature is calculated as $T=\sqrt{E^*/a}$, where $a=A/12$
is the level density parameter in the Fermi-gas model.
The yield of a particular
scission configuration with given mass and charge numbers and
deformation parameters of the fragments is proportional
to the exponential Boltzmann-factor:
\begin{eqnarray}
Y(\{A_i,Z_i,\beta_i\})\sim\exp{\left\lbrace -\frac{U(\{A_i,Z_i,\beta_i\})}{T}\right\rbrace }.
\label{yield1}
\end{eqnarray}
For given mass and charge split, the potential energy
of the dinuclear system at the scission point is a function
of deformations of the fragments, and the potential energy
surface (PES) $U_{\{A_i,Z_i\}}(\beta_L,\beta_H)$ can be drown.
Due to the Coulomb interaction between the fragments,
the deformation parameters corresponding to the minimum of PES
are larger than in the ground states of nuclei-fragments, that
indicates that the fragments at the scission point are significantly deformed.
To obtain the relative mass distribution as a function of the mass number
of one of the fragments in fission of a compound nucleus
with mass and charge numbers $A$ and $Z$, one should integrate the
expression (\ref{yield1}) over $Z_L$, $\beta_L$, and $\beta_H$, and
take into account that $A_H=A-A_L$ and $Z_H=Z-Z_L$:
\begin{eqnarray}
Y(A_L)=\frac{
\int\exp{\left\lbrace -\frac{U(\{A_i,Z_i,\beta_i\})}{T}\right\rbrace {\rm d}Z_L {\rm d}\beta_L {\rm d}\beta_H}
}
{
\int\exp{\left\lbrace -\frac{U(\{A_i,Z_i,\beta_i\})}{T}\right\rbrace {\rm d}A_L {\rm d}Z_L {\rm d}\beta_L {\rm d}\beta_H}
}.
\label{yield}
\end{eqnarray}
This distribution is normalized to unity.

We made calculations of mass distributions for five isotopes
of Hg with the mass numbers 180, 184, 188, 192, and 196. To
reduce the computation time, the calculations were restricted
only to even-even fragments since their yield is maximal
and defines the shape of the mass distribution. The inclusion of the odd-even
and odd-odd fragments can only smooth out a little the distribution
but can not appreciably change its shape.

\vspace{1cm}
\begin{figure}[here]
\includegraphics[scale=0.5]{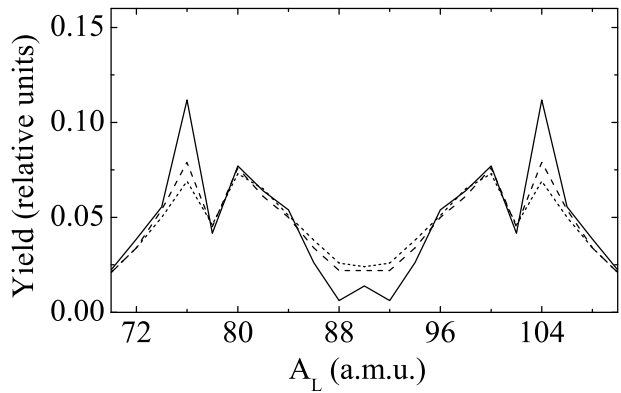}
\caption{Calculated mass distributions of fission fragments for
$\beta$-delayed fission of $^{180}$Tl (solid line), and for induced
fission of $^{180}$Hg with the impact energies of
10~MeV (dashed line) and 30~MeV (dotted line) above the Coulomb
barrier.}
\label{fig1}
\end{figure}

For $\beta$-delayed fission of $^{180}$Tl~\cite{andre}, the excitation energy
$E^*_{CN}$ of the fissioning nucleus
$^{180}$Hg does not exceed 10.44~MeV. The result of calculations
is presented in Fig.~\ref{fig1}. We obtained sharply asymmetric
mass distribution with the maximum of the yield for $^{76}$Se+$^{104}$Pd,
while the experiment gives the maximum of the yield for $^{80}$Kr+$^{100}$Ru.
Within the experimental uncertainty of 2 units in extraction
of fragment charge number this result seems to be in a rather good
quantitative agreement with the experimental data,
while qualitative agreement is excellent.
Note that for $^{80}$Kr+$^{100}$Ru we got also the maximum.

Figure~\ref{fig1} demonstrates the influence
of the excitation energy of the fissioning nucleus $^{180}$Hg
on the shape of the mass distribution. The excitation energy
reduces the shell effects and smooths out the
shape of the mass distribution. However, the influence of the
excitation energy is not so dramatic. The mass distribution
have a pronounced asymmetric shape even at the excitation
energy $E^*_{CN}$=64.2~MeV.

\begin{figure}
\includegraphics[scale=0.7]{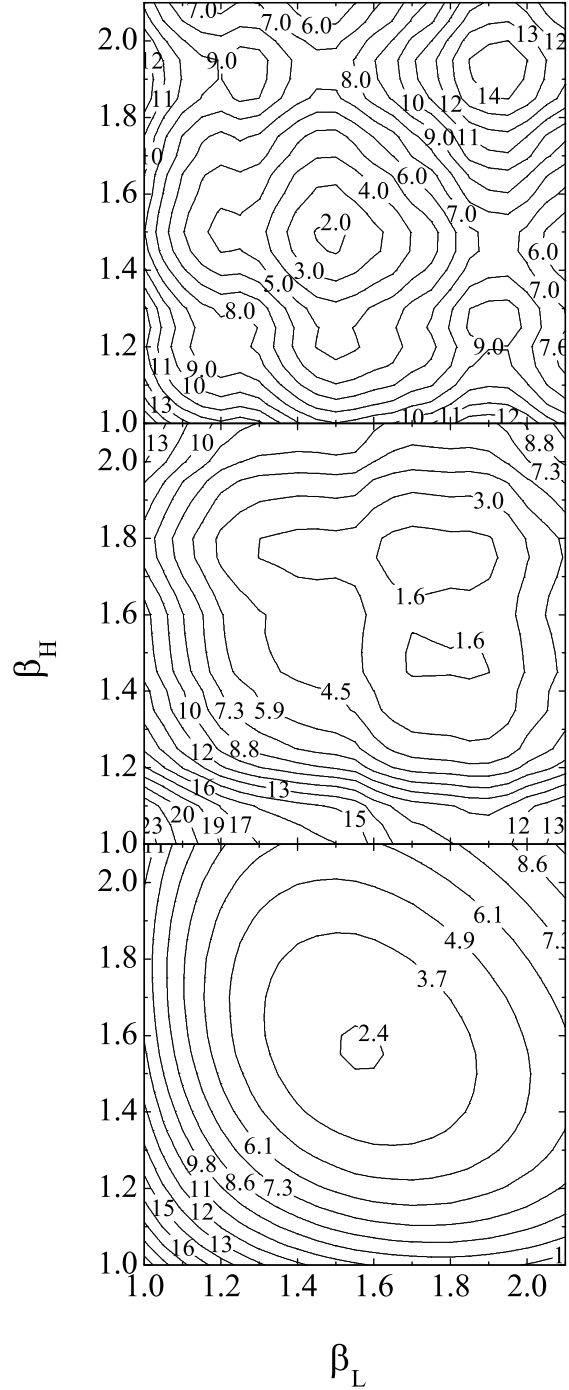}
\caption{Calculated potential energy at the scission point as a function of
deformations of the fragments in the binary systems $^{90}$Zr+$^{90}$Zr (upper part) and
$^{76}$Se+$^{104}$Pd (middle part). The energy is given in MeV
relative to the energy of the fissioning nucleus $^{180}$Hg.
For the binary system $^{90}$Zr+$^{90}$Zr,
the potential energy calculated without shell correction terms
in Eq.~(\protect{\ref{ener}}) 
is shown (bottom part).
}
\label{fig2}
\end{figure}

If one excludes the
shell correction terms from Eq.~(\ref{ener}) the PES
will have a minimum at the deformations
of the fragments of about $\beta_i$=1.6
(see Fig.~\ref{fig2}), which
are larger than the ground state deformations of the corresponding
nuclei because of the polarization effect.
Hense, the shell effects at these deformations
play crucial role for calculation of the yield, while
the shell effects at small deformations play almost
no role, since the rest part of the energy is high there.

In fission of $^{180}$Hg the symmetric scission configuration
is $^{90}$Zr+$^{90}$Zr. The shell correction for $^{90}$Zr
has a negative value $\delta U^{sh}\approx -2$~MeV near $\beta$=0,
at larger deformations it becomes positive, at $\beta$=1.6
it is equal to $\delta U^{sh}\approx 1$~MeV, then grows
further and at $\beta$=1.85 reaches $\delta U^{sh}\approx 4$~MeV.
This increases the energy of the scission configuration
at $\beta$ around 1.6 and reduces the yield of corresponding fragments.
On the contrary, the shell corrections for non-magic nuclei
in the scission configurations Kr+Ru and Se+Pd are usually
positive at small deformations
($\delta U^{sh}_L\approx 2.5$~MeV, $\delta U^{sh}_H\approx 1.5$~MeV)
and have zero or slightly negative values in the region around $\beta$=1.6
which reduces the energy of
the scission configurations and increases their contribution.
The width of the minimum of PES plays also a significant role.
Due to the integration in Eq.~(\ref{yield}), the wide minimum results
in the larger yield of corresponding fragments.
Because of the strong shell effects,
the minimum is narrow for $^{90}$Zr+$^{90}$Zr, while
for $^{76}$Se+$^{104}$Pd, where the shell effects are weaker,
it is wide (see Fig.~\ref{fig2}). This also leads to
a relative decrease of the yield of symmetric mass split.
Hence, in the fission of $^{180}$Hg we obtain the asymmetric
mass distribution with a minimum at Zr+Zr, maximum at
Se+Pd, and rather large yield of Kr+Ru.

\begin{figure}[here]
\includegraphics[scale=0.55]{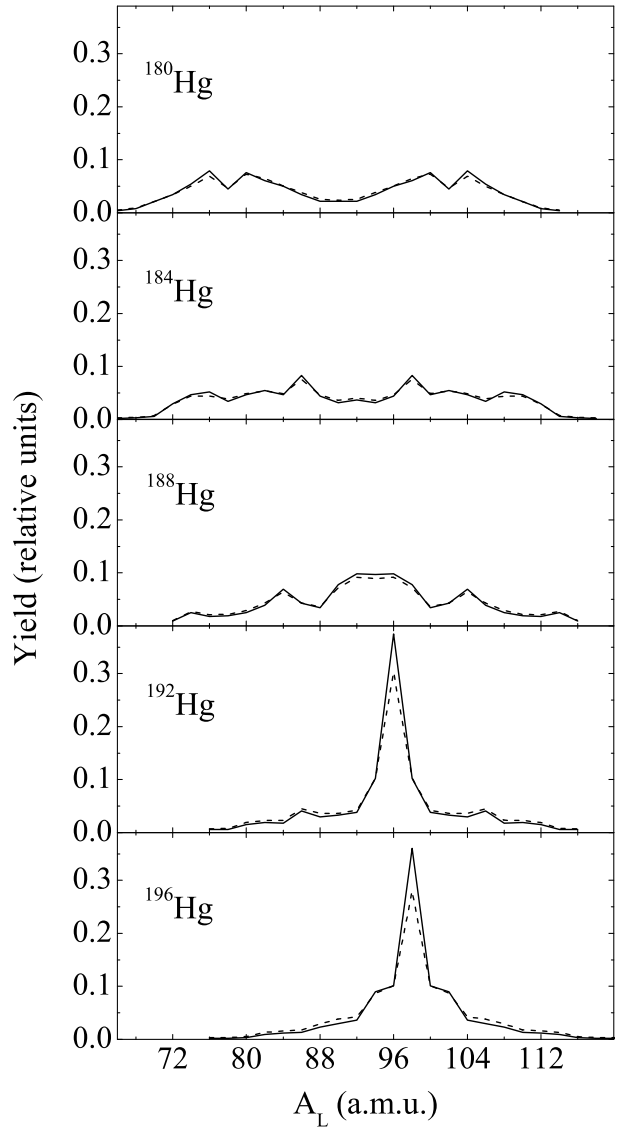}
\caption{Calculated mass distributions of fission fragments for
induced fission of $^{180}$Hg, $^{184}$Hg, $^{188}$Hg, $^{192}$Hg,
and $^{196}$Hg with the impact energies of
10~MeV (solid lines, $E^*_{CN}$=44.2~MeV, 43.9~MeV, 49.7~MeV, 62.4~MeV, 56.0~MeV for
$^{180,184,188,192,196}$Hg, respectively) and 30~MeV (dashed lines,
$E^*_{CN}$=64.2~MeV, 63.9~MeV, 69.7~MeV, 82.4~MeV, 76.0~MeV for
$^{180,184,188,192,196}$Hg, respectively) above the Coulomb barrier.}
\label{fig3}
\end{figure}

For all considered isotopes
$^{180}$Hg, $^{184}$Hg, $^{188}$Hg, $^{192}$Hg, and $^{196}$Hg
we proposed the reactions and performed the calculations of induced
fission with impact energies 10~MeV and 30~MeV above the Coulomb
barriers $V_b$:
$^{36}$Ar+$^{144}$Sm$\rightarrow^{180}$Hg ($V_b$=126.2~MeV);
$^{40}$Ar+$^{144}$Sm$\rightarrow^{184}$Hg ($V_b$=124.55~MeV);
$^{40}$Ar+$^{148}$Sm$\rightarrow^{188}$Hg ($V_b$=123.9~MeV);
$^{32}$S+$^{160}$Gd$\rightarrow^{192}$Hg ($V_b$=114.4~MeV);
$^{36}$S+$^{160}$Gd$\rightarrow^{196}$Hg ($V_b$=112.8~MeV).
Figure~\ref{fig3} shows a drastic change in the shape of the mass
distribution with increasing mass number of the
fissioning nucleus. While the mass distribution is sharply
asymmetric for $^{180}$Hg, for $^{184}$Hg the mass distribution
is rather flat, for $^{188}$Hg the mass distribution is symmetric but
very wide, and for $^{192}$Hg and $^{196}$Hg the mass distribution
have a sharply symmetric shape.

As known~\cite{goenn,hall,oganess}, the mass distribution have always
a symmetric shape if the shell effects are not taken into account.
In the present calculations the main role in the formation
of the mass distribution plays the shell effects in nuclei with
a magic neutron number $N=50$ because the symmetric
split of $^{180}$Hg consists of two magic nuclei $^{90}$Zr
while asymmetric splits of $^{180}$Hg consist of non-magic nuclei.
This marks out the isotope $^{180}$Hg and neighboring nuclei
among others. In the heavier isotopes of Hg only one fission
fragment can have the neutron number $N=50$, and hence, the
influence of the shell effects is not so strong in these nuclei
comparing to $^{180}$Hg.

With increasing mass of the fissioning nucleus, the
fragments of symmetric scission configurations
deviates from the magic $^{90}$Zr, and the role
of strong shell effects at symmetric splits decreases.
Thereby, in the heavy
isotopes of Hg the shape of the mass distribution is
generally defined by the liquid-drop part of the energy,
and we obtain a symmetric mass distribution.

\begin{figure}[here]
\includegraphics[scale=0.4]{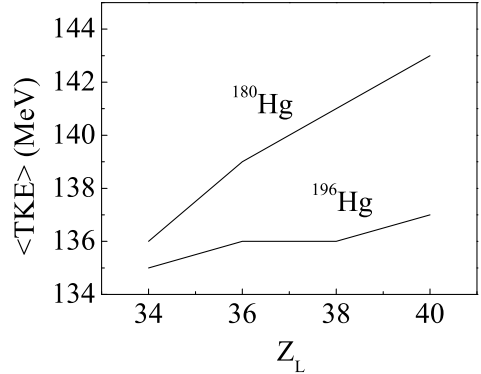}
\caption{The mean TKE of fission fragments for the
induced fission of $^{180}$Hg
and $^{196}$Hg as functions
of the mass number of light fragment.
}
\label{fig4}
\end{figure}

The scission-point model is also suitable for describing
the total kinetic energy (TKE) of the fission fragments.
We calculate the TKE supposing that all interaction
energy at the scission point transforms after fission into the
kinetic energy of the fission fragments. Therefore, the value of the
TKE strongly depends on the deformations of the fragments
at the scission point. The smaller the deformations of the
fragments, the larger the Coulomb repulsion, the larger the TKE.
The mean value of the total kinetic energy for particular binary splitting
is calculated by averaging over deformations of the fragments
on the PES:
\begin{eqnarray}
\langle TKE \rangle(\{A_i,Z_i\})=
\frac{\int[V^C(\{A_i,Z_i,\beta_i\})+V^N(\{A_i,Z_i,\beta_i\})]
\exp{\left\lbrace -\frac{U(\{A_i,Z_i,\beta_i\})}{T}\right\rbrace }{\rm d}\beta_L {\rm d}\beta_H}
{\int\exp{\left\lbrace -\frac{U(\{A_i,Z_i,\beta_i\})}{T}\right\rbrace }{\rm d}\beta_L {\rm d}\beta_H}.
\end{eqnarray}
Thus, the value of $\langle TKE \rangle$ is generally defined by the position
of the minimum of PES. In the cases of two magic
nuclei $^{90}$Zr+$^{90}$Zr and neighboring dinuclear systems
the shell corrections are strong and negative at small deformations
and grows with increasing deformations. Hence, the minimum
of the potential energy is shifted from the liquid drop
minimum $\beta$=1.6 to smaller deformations $\beta$=1.5
that leads to relative increase of $\langle TKE \rangle$.
In the scission configurations consisting of non-magic nuclei
the minimum of the potential energy correspond to
liquid-drop minimum with the deformations around $\beta$=1.6--1.65.
Figure~\ref{fig4} demonstrates the calculated dependence
of $\langle TKE \rangle$ on the mass number of the
light fission fragment for the induced fission with the impact energy
10~MeV above the Coulomb barrier. The curve rises fast for $^{180}$Hg due to
approaching to the compact symmetric scission configuration
$^{90}$Zr+$^{90}$Zr, while for $^{196}$Hg the curve is
almost horizontal since the symmetric configuration
$^{98}$Zr+$^{98}$Zr consist of non-magic nuclei. It can
be interesting to measure such dependences experimentally
and to compare with the results of our theoretical predictions.

\vspace{0.5cm}

The average value of TKE of the fission fragments can be found
by averaging over all binary systems:
\begin{eqnarray}
\overline{TKE}=
\int \langle TKE \rangle (\{A_i,Z_i\}) Y(\{A_i,Z_i\}) {\rm d}A_L {\rm d}Z_L,
\end{eqnarray}
where
\begin{eqnarray}
Y(A_i,Z_i)=\frac{
\int\exp{\left\lbrace -\frac{U(\{A_i,Z_i,\beta_i\})}{T}\right\rbrace {\rm d}\beta_L {\rm d}\beta_H}
}
{
\int\exp{\left\lbrace -\frac{U(\{A_i,Z_i,\beta_i\})}{T}\right\rbrace {\rm d}A_L {\rm d}Z_L {\rm d}\beta_L {\rm d}\beta_H}
}.
\end{eqnarray}
For $\beta$-delayed fission of $^{180}$Tl, we obtained the average TKE
of 136~MeV, that is in a good agreement with the experimental result~\cite{andre}.

The results of our calculations confirm the importance
of the shell structure in the
fission process. The account of different fragment deformations
at the scission point is necessary for the correct description
of the mass distributions and kinetic energy of the fission
fragments. Our model gives
a good description of the recent experiment, where
the asymmetric mass distribution in fission of $^{180}$Hg
was observed. This unexpected effect required a theoretical
explanation and the present work provides it. We made a
prediction of the change in the shape of the mass distribution
from asymmetric to symmetric with increasing mass number of the
fissioning Hg isotope,
and proposed the reactions to verify this prediction experimentally.

We thank Dr.~A.~N.~Andreyev for fruitful discussions.
This work was supported by DFG (Bonn) and RFBR (Moscow).


\newpage
\begin{thebibliography}{99}

\bibitem{goenn} F. G\"onnenwein, in {\it Nuclear Fission Process}, edited by C.
Wagemans (CRC Press, Boca Raton, FL, 1991).

\bibitem{hall} H. L. Hall and D. C. Hoffman, J. Radiol. Nucl. Chem. {\bf 142},
53 (1990).

\bibitem{oganess} Yu. Ts. Oganessian, J. Phys. G {\bf 34}, R165 (2007).

\bibitem{schmidt} K.-H. Schmidt {\it et al.}, Nucl. Phys. A {\bf 693}, 169 (2001); A {\bf 665},
221 (2000).

\bibitem{itkis} M. G. Itkis {\it et al.}, Sov. J. Nucl. Phys. {\bf 52}, 601 (1990); {\bf 53},
757 (1991).

\bibitem{andre} A. N. Andreyev {\it et al.}, Phys. Rev. Lett. {\bf 105}, 252502 (2010).

\bibitem{sp} B. D. Wilkins, E. P. Steinberg, R. R. Chasman, Phys. Rev.
C {\bf 14}, 1832 (1976).

\bibitem{rand} J. Randrup, P. M\"oller, and A. J. Sierk, Phys Rev. C {\bf 84}, 034613 (2011).

\bibitem{and} A. V.~Andreev {\it et al.}, Eur. Phys. J. A {\bf 22}, 51 (2004);
{\bf 26}, 327 (2005).

\bibitem{and1} A. V.~Andreev {\it et al.}, Eur. Phys. J. A {\bf 30}, 579 (2006).

\bibitem{tcsh} J. Maruhn and W. Greiner, Z. Physik {\bf 251}, 431 (1972).

\bibitem{poten} G. G.~Adamian {\it et al.}, Int. J.  Mod. Phys. E {\bf 5},
191 (1996).

\end{thebibliography}
\end{document}